\documentclass[sigconf]{acmart}
\AtBeginDocument{%
  \providecommand\BibTeX{{%
    \normalfont B\kern-0.5em{\scshape i\kern-0.25em b}\kern-0.8em\TeX}}}

\usepackage{tabularx}

\copyrightyear{2025}
\acmYear{2025}
\setcopyright{cc}
\setcctype{by}
\acmConference[ITiCSE 2025]{Proceedings of the 30th ACM Conference on
Innovation and Technology in Computer Science Education V. 1}{June 27-July 2,
2025}{Nijmegen, Netherlands}
\acmBooktitle{Proceedings of the 30th ACM Conference on Innovation and Technology in Computer Science Education V. 1 (ITiCSE 2025), June 27-July 2, 2025, Nijmegen, Netherlands}
\acmISBN{979-8-4007-1567-9/2025/06}
\acmDOI{10.1145/3724363.3729024}

\makeatletter
\def\@ACM@copyright@check@cc{}
\makeatother
\setcopyright{cc}

\begin{document}

\title[How Generative AI is Eroding Social Interactions and Student Learning Communities]{``All Roads Lead to ChatGPT'': How Generative AI is Eroding Social Interactions and Student Learning Communities}


\author{Irene Hou}
\affiliation{%
  \institution{Temple University}
  \city{Philadelphia, PA}
  \country{USA}}
\email{ihou@ucsd.edu}
\orcid{0009-0008-0511-7685}

\author{Owen Man}
\affiliation{%
  \institution{Temple University}
  \city{Philadelphia, PA}
  \country{USA}}
\email{owen.man@temple.edu}
\orcid{0009-0003-0527-1395}

\author{Kate Hamilton}
\affiliation{%
  \institution{Temple University}
  \city{Philadelphia, PA}
  \country{USA}}
\email{kate.hamilton@temple.edu}
\orcid{0009-0006-7684-2871}

\author{Srishty Muthusekaran}
\affiliation{%
  \institution{Temple University}
  \city{Philadelphia, PA}
  \country{USA}}
\email{srishty.muthusekaran@temple.edu}
\orcid{0009-0007-5806-425X}

\author{Jeffin Johnykutty}
\affiliation{%
  \institution{Temple University}
  \city{Philadelphia, PA}
  \country{USA}}
\email{jeffin.johnykutty@temple.edu}
\orcid{0009-0007-0801-6460}

\author{Leili Zadeh}
\affiliation{%
  \institution{Temple University}
  \city{Philadelphia, PA}
  \country{USA}}
\email{leili.zadeh@temple.edu}
\orcid{0009-0004-9400-5212}

\author{Stephen MacNeil}
\affiliation{%
  \institution{Temple University}
  \city{Philadelphia}
  \state{PA}
  \country{USA}}
\email{stephen.macneil@temple.edu}
\orcid{0000-0003-2781-6619}

\renewcommand{\shortauthors}{Hou et al.}

\begin{abstract}

The widespread adoption of generative AI is already impacting learning and help-seeking. While the benefits of generative AI are well-understood, recent studies have also raised concerns about increased potential for cheating and negative impacts on students' metacognition and critical thinking. However, the potential impacts on social interactions, peer learning, and classroom dynamics are not yet well understood. To investigate these aspects, we conducted 17 semi-structured interviews with undergraduate computing students across seven R1 universities in North America. Our findings suggest that help-seeking requests are now often mediated by generative AI. For example, students often redirected questions from their peers to generative AI instead of providing assistance themselves, undermining peer interaction. Students also reported feeling increasingly isolated and demotivated as the social support systems they rely on begin to break down. These findings are concerning given the important role that social interactions play in students' learning and sense of belonging.

\end{abstract}







\maketitle

\section{Introduction}

During the last two years, computing students have substantially increased their use of generative AI (genAI) tools~\cite{hou2024evolving}, closing previously identified usage gaps~\cite{hou2024effects, prather2023robots}. This growth
may be explained by 
the many associated benefits, such as personalized explanations~\cite{macneil2023experiences,  bernstein2024like, leinonen2023comparing}, intelligent teaching assistants~\cite{denny2024desirable, kazemitabaar2023studying, kazemitabaar2024codeaid, liffiton2023codehelp}, and support for identifying bugs and debugging code ~\cite{macneil2023decoding, yang2024debugging}. However, problems are also being identified, such as inequitable access to these tools~\cite{zastudil2023generative, hou2024effects}, negative impacts on students' metacognition~\cite{prather2024widening}, and threats to assessment~\cite{lau2023from, savelka2023thrilled, gutierrez2024seeing, hou2023more}. 

Although cognitive, metacognitive, and ethical aspects are beginning to be understood, social aspects are still largely unexplored. Inspired by recent studies of how help-seeking behaviors are changing due to genAI~\cite{hou2024effects, sheese2024patterns}, we investigate whether and how AI affects the social dynamics of the classroom. As students turn
to these tools for help, their social interactions
with peers, instructors, and broader learning communities are likely
to be impacted. This is important because classrooms are not just spaces for 
individual learning; they are social communities where students support each other, and where knowledge is socially constructed~\cite{prather2022getting}. If genAI disrupts social interactions, there may be negative consequences for learning and for students' sense of belonging, a factor  that is consistently linked to academic success and retention~\cite{smith2006role, allen2008third, tinto1997classrooms}.

We investigate the following research question: 

\begin{enumerate}
    \item [\textbf{RQ:}] \textbf{What are the impacts of generative AI on peer interactions and learning communities?} 
\end{enumerate}

To investigate this question, we conducted 17 interviews with computing undergraduates (8 women, 9 men) from seven R1 universities across North America. This diverse sample of participants varied in programming experience and frequency of genAI usage. Participants first compared and contrasted their experiences receiving help from peers, instructors, and the internet with genAI tools such as ChatGPT. Participants were also asked to reflect on their peer interactions since the introduction of genAI. 

Our findings suggest that genAI tools are deeply embedded within the social dynamics of the classroom. 
\begin{itemize}

    \item \textbf{GenAI interferes with peer interactions.} Instead of interacting with their classmates, students increasingly rely on AI tools for help. Students shared how GenAI acted as a mediator in their help-seeking process, since help providers often shared genAI outputs or redirected help requests to genAI rather than providing help themselves.     
    \item \textbf{Students feel isolated, demotivated, and shameful.} Students reported feeling isolated and missed solving problems collaboratively with friends. They also experienced shame associated with their use of AI tools in the presence of peers.
\end{itemize}

These findings suggest that \textbf{genAI may have harmful impacts on peer interactions and learning communities.} Traditional peer support networks appear to be eroding, which impacted both genAI users and non-users, by \textit{reducing opportunities for collaboration, mentorship, and community building.} This also presents problems for students' motivation and sense of belonging, especially for underrepresented groups who often benefit most from peer support and engagement~\cite{mishkin2019applying, horwitz2009using}. Educators must strike a balance between carefully integrating AI while fostering and sustaining the social interactions that make learning meaningful.

{\color{black}
\section{Related Work}

Recent work suggests that the growing use of genAI tools, such as ChatGPT and GitHub Copilot, is already influencing how computing students seek help and interact with course material~\cite{hou2024effects, padiyath2024insights}. Increasingly, students report that they are relying on genAI tools instead of traditional resources like peers, instructors, or the internet~\cite{hou2024effects, hou2024evolving}. These changes have prompted extensive research investigating the benefits and challenges that these tools present in computing education~\cite{prather2024beyond, prather2023robots}. Previous studies have examined the effects of genAI tools on individual learning outcomes and metacognitive processes~\cite{prather2024widening, kazemitabaar2024codeaid, sheese2024patterns, yilmaz2023effect}, while also sounding the alarm about threats to academic integrity and the potential for over-reliance on genAI tools~\cite{lau2023from, prather2023robots, sheard2024instructor, zastudil2023generative}. These works have provided valuable insight into how individual learners are affected by these tools. However, as students increasingly turn to genAI tools for help, a deeper understanding of its impacts on social learning dynamics within computing education learning communities is needed.

One key component of learning, help-seeking, is often fraught with challenges for students, who may encounter socio-emotional barriers~\cite{foong2017online} and decision-making challenges related to identifying and effectively using the appropriate resources~\cite{cheng2011an, aleven2003help}. Students want to avoid burdening their peers, they may be worried about appearing incompetent, or they may fear being rejected when requesting help. All of these factors can reduce their willingness to seek help from peers and instructors~\cite{karabenick2003seeking}. Moreover, although knowledge gained through social interactions can be invaluable, students may perceive it as coming with a social cost~\cite{chiu2006understanding}. These barriers influence how and why students decide to seek help, the types of resources they use, and when they choose to engage with peers, instructors, or the internet (e.g. internet search, StackOverflow, YouTube, etc.)~\cite{foong2017online, wirtz2018resource, price2017isnap, karabenick2003seeking, newman1990childrens}. With the emergence of genAI, prior work has shown that students increasingly prefer genAI because it lowers many of these help-seeking barriers, addressing fears of being burdensome or appearing foolish~\cite{hou2024effects}. Unlike peers or instructors, genAI tools are accessible anytime and anywhere, effectively removing barriers that have historically hindered help-seeking~\cite{doebling2021patterns}. With genAI usage also linked to perceptions of peer usage, some students maybe be more affected by these changes than others~\cite{padiyath2024insights}.

Given the social nature of help-seeking, research is needed to understand whether and how these changes affect peer interactions, relationships between students, or learning communities. Previous research consistently shows the importance of collaboration, group work, and mentorship in promoting equitable access~\cite{latulipe2018evolving, cozza2011bridging, horwitz2009using, pon2017expanding}, fostering a sense of belonging~\cite{lehman2023nevertheless, giannakos2017understanding, rosson2011orientation}, supporting self-regulated learning~\cite{prather2022getting, wortmann2024regulation}, and developing essential soft skills~\cite{porter2011experience, brown2009emphasizing}. As genAI tools become embedded within education, it is critical to examine the potential impacts on social dynamics in the classroom. 

}

\begin{table*}[h!]
\centering
\caption{We interviewed 17 undergraduate computing students at seven R1 universities across North America. The `Years' column indicates how many years the participant has been programming so far. The majors Computer Science (CS), Information Science (IS), Graphic Design (GD), Cognitive Science (CogSci), and Interaction Design (IxD) have been abbreviated.}
\setlength{\tabcolsep}{7pt} 
\renewcommand{\arraystretch}{1.25} 
\small
\begin{tabularx}{0.95\textwidth}{lllllll} 
\toprule
\textbf{ID} & \textbf{Sex} & \textbf{University Level} & \textbf{Major} & \textbf{Years} & \textbf{Frequency} & \textbf{Usage Type} \\ 
\midrule
P1 & F & 4th-year & IS/CS & 5 & Never & Previously used GenAI, but prefers not to use it \\ 
P2 & M & 3rd-year & CS & 5 & Daily & Primary source of help \\ 
P3 & F & 1st-year & DS & 1 & Sporadic & Conceptual questions \\ 
P4 & F & 2nd-year & CE/CS & 2 & Daily & Primary source of help \\ 
P5 & F & 3rd-year (returning) & CS & 6 & Never & Never used GenAI before, prefers not to use \\ 
P6 & M & 4th-year & CS & 4 & Sporadic & Tertiary source of help \\ 
P7 & M & 1st-year & CS & 10 & Sporadic & Documentation, code-writing assistance (secondary) \\ 
P8 & F & 3rd-year & CS & <1 & Sporadic & Conceptual questions, code-writing assistance (secondary) \\ 
P9 & M & 2nd-year & IS & 2 & Sporadic & Conceptual questions, starting assignments (secondary) \\ 
P10 & M & 3rd-year & CS & 3 & Daily & Primary source of help \\ 
P11 & M & 1st-year & CS & 4 & Daily & Primary source of help \\ 
P12 & M & 4th-year (returning) & CS & 16 & Daily & Primary source of help \\ 
P13 & M & 3rd-year & CS & 4 & Daily & Primary source of help \\ 
P14 & F & 2nd-year & DS & 1 & Sporadic & Debugging (secondary) \\ 
P15 & M & 3rd-year & GD/CS & 2 & Sporadic & Code-writing assistance (tertiary) \\ 
P16 & F & 4th-year & CS & 6 & Daily & Primary source of help \\ 
P17 & F & 4th-year & CogSci, IxD & 1 & Daily & Debugging (primary) \\ 
\bottomrule
\end{tabularx}
\label{tab:participant-demographics}
\end{table*}

\section{Methodology} 

To understand impacts of genAI on computing students' social interactions, we conducted semi-structured interviews with 17 computing students across 7 R1 universities in North America. Each interview lasted 30-45 minutes. We recorded the interviews via Zoom with verbal consent, and participants were discouraged from sharing their screens or videos to protect their privacy. The research was approved by our university's Institutional Review Board (IRB). 

\subsection{Participant Recruitment}

To ensure a diverse sample, we recruited participants from multiple universities through announcements made by CS faculty and within computing-related student organizations. Advertisements were also posted on relevant university subreddits and student Discord servers. Each participant was compensated with a \$10 gift card. The interviews were conducted in 2024 between June and October. Participants were all native English speakers. Further demographic information and genAI usage habits are summarized in Table \ref{tab:participant-demographics}.

\subsection{Interview Protocol and Rationale}

Interviews were semi-structured to provide flexibility in probing further into emerging themes. Participants were first asked demographic questions about their major, year in university, programming experience, and how they use genAI tools. To ground participants' perspectives to their actual experiences, we asked them about their help-seeking process. Similar to prior work on help-seeking~\cite{doebling2021patterns, hou2023more}, students ranked help resources based on their usage and trust (e.g. peers, instructors, TAs, course discussion forums, genAI, internet resources). We then asked participants to discuss the pros and cons of using genAI resources versus resources like instructors and peers. 
Participants also compared their experiences with genAI versus these other resources. The use of compare and contrast questions elicited deeper responses as participants naturally incorporated examples to justify their reasoning.

We also asked participants to reflect on whether and how access to genAI tools affected their social interactions with peers. They were then asked to share observations about any changes they noticed among their peers or within their learning communities. Asking participants to share their observations had two purposes: 1) it mitigated potential biases by encouraging descriptive rather than purely evaluative responses, and 2) it allowed interviewers to probe at complex social dynamics and potential implicit biases. 

\subsection{Thematic Analysis of Interview Transcripts}

Two researchers conducted the interviews on Zoom, which automatically transcribed the interview recordings. The transcripts were reviewed, corrected for transcription errors, and anonymized. The transcripts were then analyzed using a reflexive thematic analysis~\cite{braun2019reflecting}. Three researchers first worked individually to open-code the responses~\cite{strauss2004open}, developing their own interpretations and ensuring reflexivity. After this individual phase, researchers held periodic group discussions to share and reflect on their insights. The purpose of these discussions was to deepen their interpretation, but not necessarily to form consensus, as that is not the goal of inductive analysis~\cite{braun2019reflecting}. Rather than compromising reflexivity, the discussions supported it by encouraging researchers to interrogate their assumptions and consider alternative perspectives. Themes were developed iteratively, and each theme is presented with quotes from participants to provide interpretive context.

\section{Results}

Table \ref{tab:participant-demographics} summarizes the demographics of the participants. The participants varied by gender (8 women, 9 men), university levels (3 first-year, 3 second-year, 6 third-year, and 5 fourth-year students. This included returning students who took gap years), computing majors, and years of programming experience. GenAI usage patterns also varied: some students used it daily as their primary source of help, others used it more sporadically, and some avoided using it altogether. These varied usage patterns informed our understanding of the changing peer help-seeking interactions.

\subsection{Peer-to-Peer Relationships}

\subsubsection{GenAI as an Intermediary in Help-Seeking Interactions}

When asked about their experiences seeking help from peers, most students (13 out of 17) described how help-seeking interactions were now often mediated by genAI tools, regardless of whether they personally used these tools. 

For example, P5, a self identified \textit{non-user}~\cite{baumer2015importance} of genAI, described beginning to notice that her friends would share AI-generated code with her anytime that she asked for a `push in the right direction.' She went on to explain: 

\begin{quote}
    \textit{``Every sentence you hear: `Oh, GPT!' \textbf{Even if I don't use it, I definitely still indirectly use it. You can't really escape that}...like if I asked for help, and the help came from a human, well, they probably they got it from ChatGPT still. \textbf{They don't redirect me to GPT. They just give me what they got out of GPT}...which is why I say like, even though I haven't personally used it, I feel it's inevitable.''} (P5)    
\end{quote} 

P5's experience illustrates a shift in help-seeking from authentic peer interactions to an AI-mediated exchange. Such mediated interactions were prevalent across participants, spanning both regular and sporadic genAI users, suggesting that this phenomenon is not exclusive to non-users. From the perspective of P5, a student who was providing help, these `referrals' to ChatGPT are typical: 

\begin{quote}\textit{``Sometimes, they [peers] would ask me a question, and \textbf{I would ChatGPT it and give it back.} They're like, `Thank you, you helped me so much!' I'm like, `I did nothing.' It's such a thing now.''} (P16)\end{quote}

These `referrals' to genAI, while efficient, appeared to erode opportunities for meaningful peer interaction. For some students, this shift appeared to cause harm. P3, reflecting on a time when she expressed vulnerability, shared: 

\begin{quote}
\textit{``If you say that you're struggling, someone probably will respond, being like, \textbf{`Oh, just ChatGPT that instead.'} And that's like the biggest change I've seen.''} 
\end{quote}

{\color{black}
Students, like P3, who ask for help from peers and are rejected or redirected may be more reluctant to ask for help from their peers and friends in the future, especially given the pre-existing socio-emotional barriers for help-seekers~\cite{foong2017online}. 

Descriptions of these redirections and missed opportunities for authentic peer interaction were common in the interviews, especially among more senior students (P3, P5, P6, P10, P13, P16), who described noticing a shift that has occurred with widespread genAI use. P13 lamented this as a loss but also acknowledged the trade-offs, sharing that the \textit{``sense of comfort, knowing that my friend will be able to help me...like that camaraderie because you know you're both suffering in the assignment. [Now] most of the time, if GPT has been able to solve it, then we're not gonna ask.''} P13 elaborated by saying the perceived cost of asking a friend, whether it be time, social capital, or effort, was often no longer worth paying given the convenience of genAI alternatives, despite finding it more comforting and emotionally fulfilling to receive help from their friends. 

P5, a student who had left her R1 university for two years before returning, described the prevalence of AI tools as a \textit{`culture shock,'} observing that \textit{``[unlike] how it was a few years ago, \textbf{all roads lead to GPT.}''} This reflects a broader trend among participants, illustrated by the use of adjectives like `unavoidable' (P2) and `inevitable' (P5) to describe the mediated help-seeking interactions between peers and genAI. The use of this language suggests that these types of interaction may have rapidly evolved into an acceptable, and perhaps even expected, norm in help-seeking.

}

\subsubsection{Shame and Stigma Surrounding GenAI Usage}

{\color{black}

Despite the normalization and widespread adoption of genAI, their use is not without tension. Seven participants expressed experiencing shame or stigma associated with genAI usage. These concerns emerged unprompted, suggesting they may be highly salient aspects of students' lived experiences. Students indicated that openly using genAI---or being perceived as overly reliant on it---carried social risks, often tied to broader perceptions of academic integrity and competence. 

Students shared fears of being judged as \textbf{`lazy,'} \textbf{`stupid,'} or \textbf{`foolish'} (P4, P15, P16), and skepticism toward genAI users was common, with some describing reliance on these tools as a marker of being `less intelligent' (P14, P16). P4 and P14 recounted how these social risks were compounded by fears regarding the use of genAI in the presence of professors or authority figures, even with explicit permission to use them. For example, P4 recounted a seminar where students avoided using genAI, despite its permitted use, out of fear of being judged or accused of cheating:

}

\begin{quote} \textit{``Half the people are kind of scared. They don't want to use [ChatGPT] in class like they'll use it at home, because [at home] no one's watching them, no one cares...
\textbf{People were scared to use AI because they didn't wanna be looked down on or make it seem like they were cheating}. But to be honest, the first few people that figured it out were using Gemini.''} (P4)
\end{quote}

{\color{black}
This reluctance to engage with genAI in public reflects a new social norm students where private reliance coexists with public hesitation. P14 shared the following related perspectives, \textit{``People definitely use it. They just don't talk about it...\textbf{[Professors] allow you to use it. It still feels like it's wrong somehow.}''} 

The role of social context in mitigating shame is also evident. P15 contrasted using genAI in front of strangers versus friends: \textit{`The stranger might look at you and \textbf{see your failure}...but with friends, you just understand [why they use genAI].'} The term `failure' here is striking, indicating that reliance on genAI may be internalized as a sign of personal or academic inadequacy, with potential implications for students' self-efficacy. However, the contrast P15 draws between strangers and friends highlights the role of trust and shared understanding in mitigating these negative emotions. This speaks to the nuanced social dynamics, where students' willingness to disclose reliance on genAI may depend on how they perceive their standing within their social groups or communities.

\subsubsection{Impacts on Diverse Perspectives}

P11 noticed that ChatGPT has made people less willing to interact: \textit{``It has made \textbf{people more lazy when it comes to learning and with each other...People are less social now} 'cause my peers will tend to not ask me or our other peers questions when they might have [before].''} However, when asked if this applied to him personally, P11 acknowledged it impacted him \textit{``only a little bit. I still ask my friends what they got for their solution.''} When prompted about why he preferred help from friends over genAI, P11 likened programming to handwriting, offering an analogy: 

}

\begin{quote}
\textit{``AI will only give you the direct best answer...which will work. But it can't give you the different style of programming that humans have. My friends will have a different style of coding than I will. \textbf{It's like handwriting, which is something AI can't replicate.} AI will only give you Times New Roman, and like, people will give you handwriting.''} (P11)
\end{quote}

Four other students (P6, P8, P10, P11) also spoke about genAI  increased homogenization and diminished discourse in their learning communities. P6 was concerned that genAI could flatten critical discourse, \textit{``\textbf{When people are more satisfied with generative AI as their main source of information, that creates less discussion, which is often needed more in schools,} because discussion is what lets people in education actually adjust to the individual.''} Although the majority of students were able to observe changes to social dynamics, only a small minority of students were able to articulate the advantages and disadvantages of these observed changes.

\subsubsection{Impacts on Mentorship}

Reliance on genAI tools may potentially hinder students' development of critical help-seeking skills and access to mentorship, resulting in a growing disconnect between novice and experienced students. 
While many students discussed the tangible benefits of genAI in addressing specific questions, fewer acknowledged the intangible benefits of seeking human assistance, such as navigating the hidden curriculum~\cite{margolis2001hidden, nakai2023uncovering}, developing socio-emotional skills, and nurturing connections with peers and mentors. For example, P4 described the ability to avoid socio-emotional aspects by interacting with genAI tools, 

\begin{quote}
    \textit{``There's a lot you have to take into account: you have to read their tone, do they look like they're in a rush...versus \textbf{with ChatGPT, you don't have to be polite.}'' (P4)}
\end{quote}

Several senior students highlighted an emerging disconnect, exacerbated by genAI, between novice and experienced students. P6, a fourth-year and a CS honor society board member, shared:

\begin{quote}
    \textit{``There's a lot less interaction between entry-level and more experienced [students]...There's this disconnect: an over-reliance on AI and not really understanding problems and not asking people who actually work in the field for help.''} (P6)
\end{quote}

{\color{black}
This anecdote illustrates the well-documented, pre-existing socio-emotional barriers that come with help-seeking. Students, who may struggle to articulate questions or accept the vulnerability that comes with asking for help, can increasingly turn to genAI to avoid these challenges. In this case, AI may be reinforcing these avoidance behaviors. As P15, a senior student, sums up: \textit{``It \textbf{seems} that GPT has everything, every answer. So you find students not then interacting with other classmates or colleagues.''}

However, multiple students recognized the role instructors and peers have in helping them navigate the hidden curriculum~\cite{margolis2001hidden}. P9 describes this value of finding information that you might not know to look for when interacting with peers and instructors: 
}

\begin{quote}
    \textit{``Human conversations can have the added benefit of, like, \textbf{you can get knowledge that you weren't really intending to get}... Professors who really know their stuff can explain it and also connect it to different concepts. I don't think ChatGPT can do that.''} (P9) 

\end{quote} 

\subsubsection{Impacts on Motivation}

According to students, peers provided unique value by inspiring and motivating them. For example, students described how engaging with peers exposed them to relevant opportunities (P2, P4), fueled their passion for computing (P6, P7, P15, P17), and helped them stay engaged while learning (P13, P15). P17 said that connecting with humans and sharing life experiences inspired their interest in computing: 

\begin{quote}
   \textit{``\textbf{[My classmates'] experiences can be shared, their feelings, whatever desires they have, what drives them - it can also impact me.} Like, `That was pretty cool, I kind of want to go into that, too'... I had a friend interested in designing a product for image generation AI systems, and I just saw their passion.\textbf{ Being passionate about it made it so interesting to me.}''} (P17)
\end{quote}

Students also spoke about how friends played an important role in keeping them engaged and motivated. P13 explained that, \textit{``When we're here with friends, there's more of the social aspect, which makes it more fun. Whereas with GPT, you're like, `Okay, well, this is another tool.'''} Similarly, P15 explains how shared accomplishments and working hard together was a major motivator:

\begin{quote}\textit{``With friends, when you get it right together, it feels like an achievement; \textbf{it's you and your friends grinding it out.} I'm more motivated with my friends than GPT.''} 
\end{quote}
In contrast, P11, a self-taught CS student, admitted that genAI made him \textit{``very unmotivated to learn programming, which is something I used to really care about. I feel like, what point is there to it anymore, since AI can do it so well."} Unlike P15, whose motivation was tied to peer collaboration, P11's motivation was tied to developing his own skills, which he felt had lost meaning due to genAI. For students who are motivated primarily by competence or mastery, genAI may make learning feel less meaningful, while students motivated by peer collaboration may be impacted differently.

Those who said genAI tools increased their motivation described reduced wheel-spinning (P12, P16) and the ability to explore topics more easily (P9, P12). For instance, P12 said, \textit{``Rather than spend a whole evening stuck on a problem, I can quickly identify the issues and...make a lot more progress, and then I spend less time frustrated and more time programming because I enjoy it."} 

While genAI tools can provide significant benefits when it comes to solving the tangible problems that students face in their assignments and work, there are aspects that cannot be replaced by these tools. The drive, unique interests, and passions of one's peers and community cannot be replicated by genAI tools. 

\subsubsection{Impacts on Community and Feelings of Isolation}

Towards the end of the interview, participants were asked how genAI usage may or may not be affecting their learning community and if they had any of their own experiences on the matter. 
The majority of students (11 out of 17) 
had noticed impacts to their community, such as increased feelings of isolation.

P2, a daily user of genAI, shared how genAI tools made it easier for him to become more isolated: \textit{``I don't really actively go out of my way to socialize with people... \textbf{So if I'm relying more on GPT, I might be more isolated in my room, instead of having to go out and talk to people.}''} P2 later observed how personal experience and insight from his peers was an important part of developing a better understanding of the field and finding future career opportunities: \textit{``If you're alone, you might not even know about what's out there, how to bolster your resume, things like that.''} However, this awareness did not appear to reduce his genAI reliance or feelings of isolation.

In addition, students observed that activity in online communication platforms like Discord was decreasing with the rise of genAI. As these crucial community spaces become less active, students are cut off from a source of social support. P16 highlights this problem: 

\begin{quote}
    \textit{```We used to in every class have a Discord. It used to be like a lot of people just asking questions about maybe like, a lab or a homework... I guess everyone's just ChatGPT now. Like the new classes that I have now, \textbf{we still have the Discord, but nobody really talks} because most or all the questions are answered by ChatGPT.''}
\end{quote}

P17, a student who no longer used Discord, shared a theory about why this is happening: \textit{``I did browse a lot more on like Discord and Slack [before genAI] for what other people asked...once I started using ChatGPT a bit more, I stopped browsing through Slack and Piazza.''}

Students' responses suggest feelings of isolation that are compounded by the erosion of social spaces on which they once relied. This raises concerns that learning communities may be at risk as students disengage from them.

\section{Discussion}

Our findings suggest that genAI tools may be reshaping the \textit{social fabric} of computing education. Students described that many of their peer interactions are now frequently mediated by genAI. For example, students reported that help requests were often redirected to genAI or included genAI outputs instead of direct support from peers, a trend that even affected non-users of genAI~\cite{baumer2015importance}. This mediation undermined the value of peer interactions, and students unanimously perceived a reduction in peer interactions as students receive help from genAI instead of their classmates. Traditionally, peer interactions fostered camaraderie and mutual support which contributed to the formation of informal student learning communities~\cite{lave2001legitimate,bandura1977social}. However, as genAI disrupts these social interactions, the mechanisms that drive community building may be eroding.


Older students also shared concerns that incoming students are becoming less connected to senior mentors. This loss of mentorship opportunities reduces access to the hidden curriculum (i.e.: unwritten rules, strategies, and cultural norms that are essential for success~\cite{margolis2001hidden}). Informal opportunities for interaction can serve as entry points into learning communities through legitimate peripheral participation~\cite{lave2001legitimate}, and this reduced access will disproportionately impact first-generation students, who can not rely on their family to help them navigate the hidden curriculum~\cite{jack2016no}.  

Reductions in peer interactions and mentorship appear to have emotional consequences. Many students reported feeling lonely; some described how their passion for computing was sparked and sustained through collaboration and commiseration with peers. In contrast, genAI tools improve efficiency but cannot replace a friend who provides that socio-emotional or motivational support. 

In addition to these social and emotional harms, our findings suggest that recently discovered metacognitive pitfalls associated with AI usage~\cite{prather2024widening} such as being misled by suggestions, may be further exacerbated. Students often rely on on socially shared regulation to scaffold their own self-regulation abilities by observing how their peers set goals, monitor progress, and adjust their strategies~\cite{hadwin2011self, schunk1997social}. Without this scaffolding, students must face these new metacognitive challenges with even less less support.

Our findings are both surprising and troubling. Students' computing identities are socially constructed~\cite{lunn2021educational}, they lean on each other for socio-emotional support and motivation~\cite{prather2022getting, davis2023equitable}, and they develop a sense of belonging, which has been consistently linked to retention~\cite{tinto1997classrooms}. If genAI is interfering with these social help-seeking processes to the extent we observed through these interviews, research is urgently needed to protect these critical social aspects of our learning environments. The social fabric of our learning communities---the peer interaction and connection that is integral to student success---appears to be at risk as genAI alters how students seek help and collaborate. Given the growing use of genAI~\cite{hou2024effects, hou2024evolving}, researchers and educators must be increasingly mindful about fostering healthy peer interactions and scaffolding the development of formal and informal learning communities.

{\color{black}

\subsection{Limitations} 
This study focuses on the perspectives of undergraduate computing students and the perceived impacts of genAI on their peer interactions and learning communities. However, the changes described by participants are anecdotal and have not yet been corroborated empirically. Our sample size is small and focuses on the perspectives of students in North America. Consequently, our findings should not be generalized to all cultural contexts, as social interactions can differ across cultures.  
This study does not aim to generalize but to generate critical early insights into a changing social landscape, for which interviews are an appropriate method. Future work is needed to confirm the trends observed in this work. 

\section{Conclusion}

In this paper, we conducted interviews with 17 students from multiple universities to investigate the effects of genAI on social dynamics. Although exploratory in nature, our findings reveal concerning trends such as reduced social interactions, missed opportunities for mentorship, diminished motivation, and feelings of isolation. Despite some benefits, AI may be inadvertently destroying the very \textit{social fabric} that supports meaningful learning. Going forward, it is necessary to balance the integration of AI with the irreplaceable value of human interaction.

}

\balance
\bibliographystyle{ACM-Reference-Format}
\bibliography{sample}


\end{document}